\documentclass[
]{ceurart}

\usepackage{listings}
\usepackage{acronym}

\newacro{ADHD}[ADHD]{Attention Deficit Hyperactivity Disorder}
\newacro{DHI}[DHI]{Digital Health Interventions}
\newacro{AI}[AI]{Artificial Intelligence}
\newacro{UI}[UI]{user interface}
\newacro{GUI}[GUI]{graphical user interface}
\newacro{HCI}[HCI]{Human-Computer Interaction}
\newacro{UX}[UX]{user experience}
\newacro{HRC}[HRC]{Human-Robot Collaboration}
\newacro{HRI}[HRI]{Human-Robot Interaction}
\newacro{SUS}[SUS]{system usability scale}
\newacro{CSCW}[CSCW]{computer-supported cooperative work}
\newacro{MR}[MR]{Mixed Reality}
\newacro{CVE}[CVE]{Collaborative Virtual Environment}
\newacro{AR}[AR]{Augmented Reality}
\newacro{AV}[AV]{Augmented Virtuality}
\newacro{VR}[VR]{Virtual Reality}
\newacro{PRISMA}[PRISMA]{Preferred Reporting Items for Systematic Reviews}
\newacro{PRISMA-Scope}[PRISMA-ScR]{Meta-Analyses Extension for Scoping Reviews}
\newacro{eHMIs}[eHMIs]{external Human-machine interfaces}
\newacro{SAR}[SAR]{Socially Assistive Robots}
\newacro{OCD}[OCD]{Obsessive Compulsive Disorder}
\begin{document}

\copyrightyear{2026}
\copyrightclause{Copyright © 2026 for this paper by its authors. Use permitted under Creative Commons License Attribution 4.0 International (CC BY 4.0).}

\conference{CHI '26: Co-Living: Reality <-> Virtuality,
  April 13--17,
  2026, Barcelona, Spain}

\title{A Roadmap of Mixed Reality Body Doubling for Adults with ADHD}

\author[1]{Valerie Tan}[%
orcid=0009-0005-8159-4027,
email=valerie.tan@tu-dortmund.de,
]
\cormark[1]
\address[1]{TU Dortmund University,
  August-Schmidt-Straße 1, 44227 Dortmund, Germany}

\author[1]{Kimberly Hegemann}[%
orcid=0009-0001-4608-951X,
email=kimberly.hegemann@tu-dortmund.de,
]

\author[1]{Jens Gerken}[%
orcid=0000-0002-0634-3931,
email=jens.gerken@tu-dortmund.de,
]

\cortext[1]{Corresponding author.}

\begin{abstract}
Adults with \ac{ADHD} may use a self-management technique known as Body Doubling, in which the participant employs the presence of one or more agents as a means of initiating and completing tasks. We developed a framework on body doubling with twelve dimensions to better understand the characteristics of body doubling and discover future research directions for developing and testing body doubling for adults with \ac{ADHD}. Our framework accounts for individual motivation, agent-related dimensions, interaction related dimensions, contextual dimensions, and efficacy. These dimensions show existing research gaps such as limited mixed reality prototypes, possibilities for more interactive body doubles, and the need for empirical studies to further understand of body doubling and adults with \ac{ADHD}. 
\end{abstract}


\maketitle

\section{Introduction}
Many adults today manage \ac{ADHD} or impairments highly associated with it, with estimates around  2-4\% of the adult population \cite{Weibel2020-sa} and possible underdiagnosis \cite{Ginsberg2014-pn}. 
\ac{ADHD} is considered a neurodevelopmental disorder characterized by persistent inattention, hyperactivity, and impulsivity \cite{icd11_2025, Weibel2020-sa, Sjowall2013-yo}. 
Adults with \ac{ADHD} grapple with functional impairments that impact their daily life \cite{Holst2019}. They include negative impacts in work life, education, social relations, and living situations \cite{Gjervan2012-xe, Holst2019}. 

Common approaches for managing \ac{ADHD} include pharmacological treatments and psychotherapy \cite{Weibel2020-sa, Ginsberg2014-pn}. 
Recent research on neurodivergence, including \ac{ADHD} \cite{Spiel2022-ea, Eagle2024-cb} suggest motivations for additional management approaches that go beyond systems such as planners, calendars, to-do lists, or site blockers \cite{Moell2015-az}.

One technique that adults with \ac{ADHD} use is body doubling, where a person does a task such as studying, working, or chores in the presence of one or more other agents, referred to as body doubles \cite{Eagle2023-tm}. Body doubling can potentially help people initiate tasks, stay focused, and complete tasks through various theorized mechanisms such as social pressure, companionship, or as a means to overcome the anxiety of doing a difficult task \cite{Eagle2023-tm, Eagle2024-cb, Lee2021-rj}. 
This technique or strategy can be done more ``traditionally'' in the physical world, such as through the presence of other people in coffee shops and libraries, or be mediated by technology \cite{Eagle2023-tm, Eagle2024-cb, Taber2019-ad}. In the latter scenario, body doubling can be done live through video calls and livestreams or through pre-recorded videos such as ``study with me'' videos, where a real person or even an animation does a task in parallel \cite{Eagle2023-tm, Eagle2024-cb, Taber2019-ad, Lee2021-rj}. One very well-known example of body doubling is the ``lofi hip hop radio - beats to relax/study to,'' which contains an animated loop of ``LoFi Girl'' studying \cite{Eagle2023-tm, Taber2019-ad, lofigirl_channel}. 
Already, the type of body doubles vary in different ways: in terms of technology, type of task, synchronicity, embodiment, and much more. Despite this variety, body doubling is a fairly new and under-researched topic in research \cite{Eagle2023-tm, Eagle2024-cb}. 
Many existing technology-mediated body doubling approaches involve two-dimensional videos \cite{Lee2021-rj}, which relies on the person using the video to create an immersive bridge between the video's setting and their everyday task embedded in the real world.
Incorporating emerging technologies such as mixed reality or robotics to create body doubles could potentially provide an opportunity for more immersive and customizable environments, especially for adults with \ac{ADHD}. 
Other dimensions and attributes of body doubling may also impact its effectiveness in improving the everyday lives of adults with \ac{ADHD}. Therefore, we created a framework on body doubling that can serve as a roadmap for further research on how we can design body doubling setups that help adults with \ac{ADHD}, and how each of the dimensions could impact the efficacy of body doubling. Although our focus during this framework development is on adults with \ac{ADHD}, our framework could potentially be adapted in other contexts and with other groups. Overall our research question is as follows: \\

\textbf{RQ:} In what ways can adults with \ac{ADHD} effectively apply body doubling to improve their everyday lives? \\

In this short paper, we will discuss related work, introduce our framework on body doubling, and explain how we can use our framework as a roadmap to determine new opportunities for body doubling prototypes and empirical studies, especially in the context of empowering adults with \ac{ADHD}.  



\section{Related Work}
\subsection{Body Doubling}
Research on body doubling in HCI has been mainly very recent and limited in number \cite{Eagle2024-cb} and similarly, HCI research on adults with \ac{ADHD} has also been limited \cite{Spiel2022-ea, Tan2024-mr, Tan2025-lr}. Notably, Eagle et al. \cite{Eagle2023-tm} proposed in their poster and in their follow-up journal article \cite{Eagle2024-cb} a two-dimensional model of body doubling for neurodivergent individuals (including people with \ac{ADHD}, Autism, \ac{OCD}, and other conditions). 
One dimension is a continuous space/time axis where one end represents same time and space and the other end represents different time and space \cite{Eagle2023-tm, Eagle2024-cb}. The other dimension is a continuous mutuality axis with one end representing ambient companionship with less awareness and mutuality, and the other end representing performance/accountability with more awareness and mutuality \cite{Eagle2023-tm, Eagle2024-cb}. 
Based on survey data, Eagle et al. (2024) further reported uses cases, types of activities, modality, location, identity of body doubles, and theorized mechanisms for body doubling \cite{Eagle2024-cb}. 

Lee et al. \cite{Lee2021-rj} explored the mechanisms behind ``study with me'' videos, which are prerecorded videos or livestreams of people doing work, with limited interaction with the audience. Similarly to Eagle et al. \cite{Eagle2024-cb, Eagle2023-tm} we consider this to be a specific digital form of body doubling. It is important to note however that the authors did not focus on neurodivergence.   

\subsection{Specific Applications of Body Doubling}

A preprint by Ara et al. \cite{Ara2025-bv} investigated a \ac{VR} body doubling environment for adults with \ac{ADHD}, with a specific use case in construction. They conducted an empirical study where participants performed a bricklaying task in \ac{VR} under three conditions: without a body double, with a human body double, and with an \ac{AI} body double \cite{Ara2025-bv}. 
This work is one of the few known empirical studies on body doubling approaches specifically for adults with \ac{ADHD}. 

\ac{SAR} have very recently been proposed as body doubles. O'Connell et al. \cite{OConnell2024} conducted a study on using a Blossom robot as a body double for university students. In their study, Blossom acts as an ambient companion and is limited to idle motions. Meanwhile, Lalwani et al. \cite{Lalwani2025} developed a socially assistive robot named Alex with not only a body double presence, but also additional ``direct, interactive assistance in time management, task prioritization, and emotional regulation'' \cite{Lalwani2025}. 

Deshmukh \cite{Deshmukh2025-pl} proposed an \ac{AI} agent and productivity assistant that helps people with \ac{ADHD} stay focused in work environments. Their body doubling component is an ``ambient presence'' and does not have a visual component but instead manifests as an voice-only assistant with additional ``interventions'' and nudges that go beyond body doubling \cite{Deshmukh2025-pl}, similar to Lalwani et al.'s more active approach.   


\section{Proposed Body Doubling Framework}
We iteratively developed a work-in-progress body doubling framework or model based on existing literature, research on \ac{ADHD} communities, and discussions between the authors. We developed our model with a strong \ac{ADHD}-specific lens throughout the process.
Our body doubling model expands upon Eagle et al.'s model by adding additional levels of their dimensions (Space/Time and Mutuality/Awareness) \cite{Eagle2024-cb} as well as additional dimensions, which Eagle et al. empirically derived and explained, but did not explicitly include in their model. In addition, our model provides a roadmap for exploring underexplored dimensions or emerging technologies such as \ac{MR} and \ac{SAR}. We will introduce all of these dimensions in the following sections. 

\begin{figure*}[htbp]
    \centering
    \includegraphics[width=0.8\linewidth]{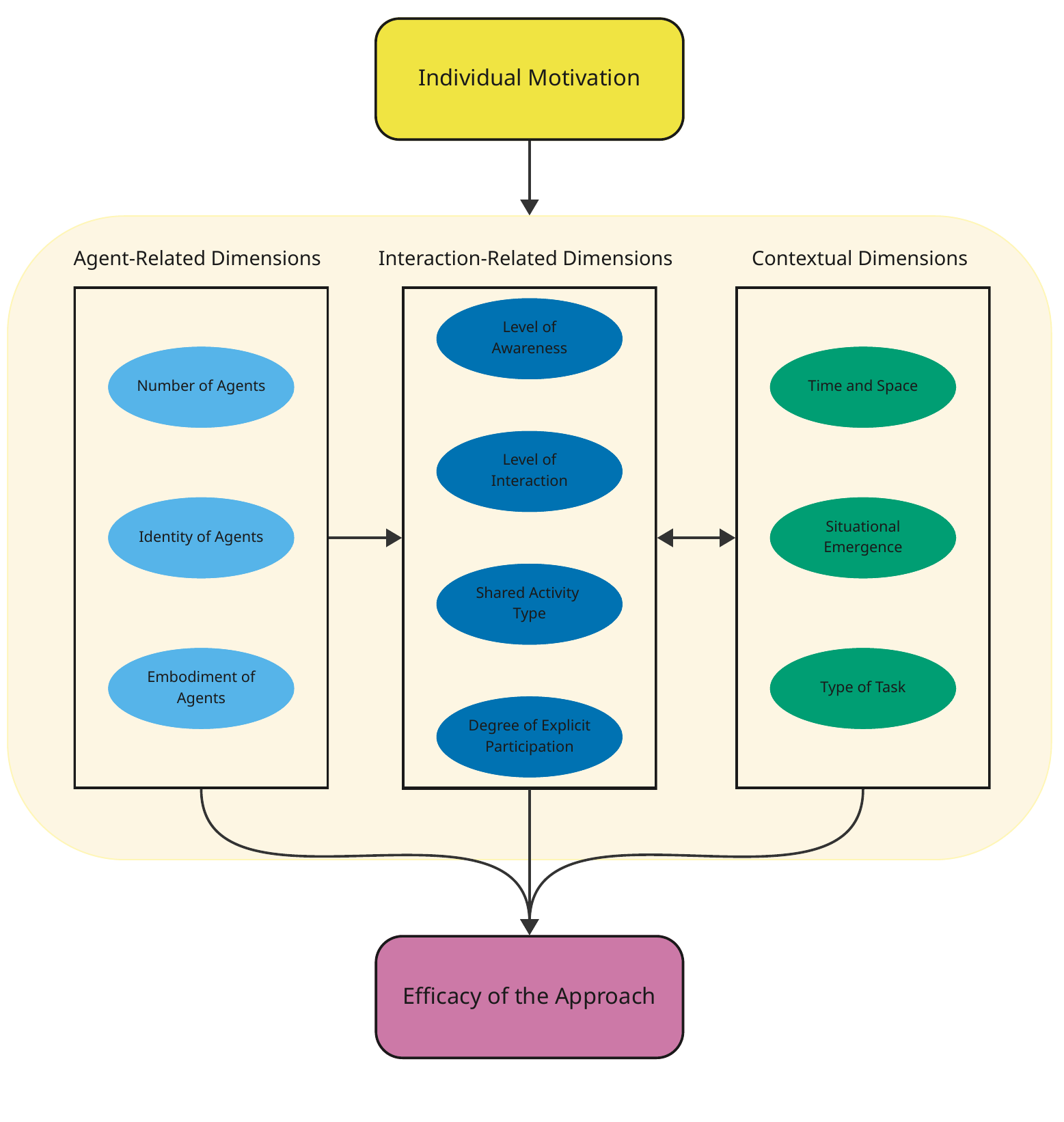}
    \caption{A work-in-progress framework or model of relevant dimensions that describe a body doubling setup.}
    \label{fig:model}
\end{figure*}

\subsection{Body Doubling Selection} 
The individual using body doubling as a technique has one or more purposes at a self-management level that they expect a body double to fulfill. 
We define the dimension \textbf{Individual Motivation} as expected self-management effects which include, but are not limited to, accountability \cite{Eagle2024-cb, annavarapu2024comparative}, task initiation \cite{Eagle2024-cb}, motivation \cite{Eagle2024-cb, Lee2021-rj}, reduced distractions \cite{Lee2021-rj}, social pressure to complete tasks \cite{Eagle2024-cb}, external structure (via reminders or timers) \cite{Eagle2024-cb, Lee2021-rj}, habit cues \cite{Lee2021-rj}, or emotional regulation support \cite{Eagle2024-cb, annavarapu2024comparative, Lee2021-rj}. 
Many of these motivations are predominant in the motivations of people with \ac{ADHD} \cite{Tan2025-lr} as well the motivations of people with other forms of neurodivergence \cite{Eagle2024-cb}. 

\subsection{Attributes of Body Doubling Setups}
The central section of our proposed model in~\autoref{fig:model} consists of dimensions pertaining to the body doubling setup itself. We further categorized the dimensions into three subgroups: Agent-Related Dimensions, Interaction-Related Dimensions, and Contextual Dimensions. 

\subsubsection{Agent-Related Dimensions}
Agent-Related Dimensions refers to attributes of the body double(s) itself. The \textbf{Number of Agents} dimension has two levels: \textit{solo with one body double} and \textit{solo with multiple body doubles} \cite{Eagle2024-cb}. 

The \textbf{Identity of Agents} dimension defines and distinguishes body doubles or agents in three ways. The first category is by \textit{familiarity status}, and thus agents such as peers or friends are \textit{known} agents, and strangers in a café or streamers are \textit{unknown}. The second category is between \textit{human} or \textit{non-human} (avatar or robots, including robots with humanoid-like design). The third category differentiates agents between \textit{living} and \textit{non-living} agents. 

The \textbf{Embodiment of Agents} dimension refers to the extent to which a body double or agent is materialized. We have three levels for this: \textit{low embodiment}, such as calm technologies including ambient light, sound, or subtle notifications (e.g., LED color shifts, soft ambient sound) or AI agents with no avatar; \textit{medium embodiment}, such as two-dimensional avatars or prerecorded video (e.g., ``study with me'' video); and \textit{high embodiment}, such as humans (e.g. a co-present person), three-dimensional avatars in AR/VR, or robots with a clear social presence (e.g. a socially responsive robot). In our framework, we consider low embodiment agents as as agents, even though they do not have a physical presence or even a visible avatar. One notable example of this is Deshmukh's body doubling AI agent \cite{Deshmukh2025-pl}.  

\subsubsection{Interaction-Related Dimensions}
The dimensions in this subgroup relate to interaction between the individual partaking in body doubling and one or more agents. 

One such relevant dimension is the \textbf{Level of Awareness}. This stems from a body doubling community on the platform Discord \cite{bodydoublingDiscordBody}, as well as from one factor of the mutuality and awareness axis in Eagle et al.'s model \cite{Eagle2024-cb}. We adapted the definition of workspace awareness proposed by Gutwin and Greenberg \cite{Gutwin2002-by}. They describe awareness as knowledge about the environment bounded in time and space that must be maintained, kept up to date and sustained through interaction with the environment. In their framework, the main goal is some task in the environment, and awareness is the secondary goal that supports the main task. 
Applied to body doubling, we differentiated awareness of the supported individual into three levels. \textit{Peripheral awareness} describes situations where the body double is perceived merely as a background presence, maintained passively and with minimal cognitive effort (for example calm technology, soft music, idle avatars). \textit{Attentive awareness} occurs when the presence of the body double can be brought into focus as needed, for example by glancing at them or briefly noticing their behavior. \textit{Continuous awareness} refers to an ongoing attentional focus, where the body double demands engagement and requires continuous perception and action, such as when directly conversing while working.

\textbf{Level of Interaction} with the body double is another dimension with three categories: \textit{high interaction}, characterized by two-way engagement and mutual feedback \cite{Eagle2024-cb}; \textit{medium interaction}, involving co-presence with implicit acknowledgment and some engagement; and \textit{low interaction}, in which the body double serves only as a passive or symbolic presence \cite{annavarapu2024comparative}. 

A third dimension, \textbf{Shared Activity Type}, describes to what extent the type of activities or tasks are similar between all agents in the body doubling setup, including the individual in focus explicitly employing body doubling. One characteristic is \textit{same task} where all agents may perform the same task (for example, cleaning or studying together). Another characteristic is \textit{different tasks} where all agents are engaged in entirely different tasks so that no two body doubles share the same task (for instance, one agent works while another studies, and any additional agents must do a distinct type of task). The third characteristic combines the first two and is described as \textit{mixed/hybrid tasks} where some agents may perform the same task while others perform different tasks. This characteristic can only apply to groups with more than two agents.

Eagle et al. \cite{Eagle2024-cb} further point out that it is not always necessary for the involved agents to knowingly participate in body doubling. For example, an individual may sit in a café and make use of the presence of strangers without their knowledge. We adapted Eagle et al.'s mutuality and awareness axis in their body doubling model. In particular, we focused on the mutuality part in this dimension. 
Thus, the dimension \textbf{Degree of Explicit Participation} can be specified in three levels: \textit{reciprocal participation} in which all agents explicitly consent to and intentionally engage in body doubling; \textit{unilateral participation} where only the recipient deliberately engages in body doubling, while all other agents are present but unaware of their role as a body double (e.g., strangers in a café or an avatar) and \textit{mixed participation}, in which some agents knowingly engage in body doubling, whereas others do so unknowingly (e.g., two friends meet in a library to body double and additionally utilize the not explicitly participating visitors of the library as body doubles).

\subsubsection{Contextual Dimensions}
The Contextual Dimensions refer to dimensions that as the name suggests, describe the context of the body doubling setup or involved tasks. 

\begin{figure*}[htbp]
    \centering
    \includegraphics[width=0.5\linewidth]{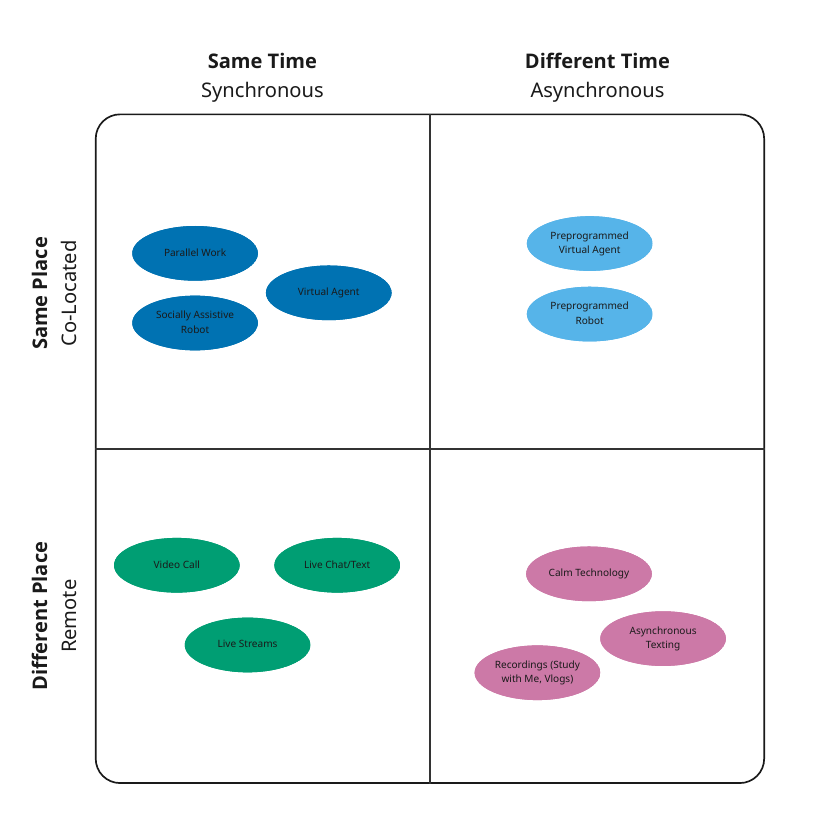}
    \caption{Application of Johansen's Time-Space matrix in the context of body doubling.}
    \label{fig:time_space}
\end{figure*}

The \textbf{Time and Space} dimension respectively refers to whether the body double(s) and the supported person are respectively in the same place and/or doing tasks synchronously \cite{annavarapu2024comparative, Eagle2024-cb}. We adapted Johansen’s time-space matrix as shown in ~\autoref{fig:time_space}\cite{johansen1988groupware, PENICHET2007237}. 
Here, time is differentiated as \textit{synchronous/same time} and \textit{asynchronous/different time}, while space is differentiated into \textit{co-located/same place} and \textit{remote/different place}. 
In the quadrant ``same time, same place'' we find co-located synchronous body doubling, which may include parallel work, the presence of an immersive mixed reality body doubling agent, or social robots. 
In the quadrant ``same time, different place'' we encounter remote synchronous body doubling, which encompasses live streams such as ``study with me'' videos \cite{Jhuang_2022, Lee2021-rj} or the Lofi Girl stream \cite{Taber2019-ad}, as well as live video calls or live texting. 
The quadrant ``different time, different place'' corresponds to remote asynchronous body doubling and includes prerecorded ''study with me'' content. 
Finally, in the quadrant ``different time, same place,'' we find co-located asynchronous body doubling, for example with pre-programmed virtual agents or robots.
Eagle et al. \cite{Eagle2024-cb} include two of the quadrants, ``same time, same place'' as well as ``different time, different place,'' and our model expands on that by allowing further differentiation between setups such as live streamed ``study with me'' videos (``same time, different place'') and recorded ``study with me'' videos (``different time, different place''). 

We can further characterize body doubling setups based on how they emerge, which refer to as a dimension called \textbf{Situational Emergence}. At the \textit{spontaneous/emergent} level, we find that body doubling may emerge spontaneously through unplanned co-presence, for instance when individuals find themselves working in the same room and unintentionally benefit from each other's presence \cite{Eagle2024-cb, addADHDBody}. 
At the \textit{informally co-created} level, it can also be initiated informally by peers who deliberately arrange to work alongside one another without relying on any formal structure, as described in community-organized body doubling sessions or casually arranged study meetings \cite{annavarapu2024comparative, Eagle2024-cb}. 
At the final level, \textit{formally organized}, body doubling takes place within more structured or institutionalized contexts such an \ac{ADHD} coaching setting, support groups or guided co-working programs \cite{addADHDBody}.

The \textbf{Type of Task} is also relevant in defining body doubling \cite{annavarapu2024comparative, Eagle2024-cb}. For this dimension, we applied the International Classification of Activities for Time Use Statistics (ICATUS). The activities in this classification span a wide range, including \textit{employment and related activities}; \textit{production of goods for own final use}; \textit{unpaid domestic services for household and family members}; \textit{unpaid caregiving services for household and family members}; \textit{unpaid volunteer, trainee, and other unpaid work}; \textit{learning}; \textit{socializing}, \textit{communication}, \textit{community participation, and religious practice}; \textit{cultural, leisure, mass media, and sports activities}; and\textit{ self-care and maintenance}. These activities serve as the basis for categorizing the types of tasks that can be carried out with the support of body doubling.

\subsubsection{Relationships between Dimensions}
We theorized relationships between subgroups, since one or more dimensions in one subgroup may impact one or more dimensions in another subgroup. 
For example, the Embodiment of Agents and Identity of Agents dimensions may impact the Level of Interaction in a setup. An example of this is a body doubling setup between two friends which leads to a high level of interaction. Therefore, we theorized that Agent-Related Dimensions would impact Interaction-Related Dimensions as shown in~\autoref{fig:model}. Similarly, we theorized that Contextual Dimensions would impact Interaction-Related Dimensions, as it is possible that asynchronous and remote settings may impact the Level of Interaction. We also theorize that Interaction-Related Dimensions would impact Contextual Dimensions, since interaction dynamics can influence the type of task that the agents do.   

We chose to simplify our model by not theorizing direct relationships between dimensions. Otherwise, given 10 distinct dimensions, we would need to consider all possible bidirectional edges in a complete bigraph \cite{weisstein_completedigraph} with 10 nodes, so therefore in our case, there would be a maximum of 90 possible relations.  

\subsection{Efficacy of Body Doubling Setups}
Measuring how well a body doubling system is effective can be done in a variety of ways. They range from qualitative approaches \cite{Eagle2024-cb, Eagle2023-tm, Lalwani2025, OConnell2024, Ara2025-bv} to quantitative scales related to usability \cite{Lalwani2025}, perceived improvement in functioning \cite{OConnell2024, Ara2025-bv}, or task performance \cite{Ara2025-bv}.
Future work should evaluate various body doubling setups that vary by selected dimensions as described previously. 
There should also be more empirical validation of body doubling setups with people with \ac{ADHD}. 


\section{Discussion}
We can use our model as a roadmap to determine further areas of research. The first direction entails opportunities for prototyping and testing in mixed reality. The second direction extends beyond body doubling and incorporates more interactive systems. 
The final direction suggests further empirical studies on specific user groups, such as adults with \ac{ADHD}, or more studies on the dimensions of body doubling. 
  
\subsection{Opportunities for Mixed Reality}
Because the availability of virtual agents with high embodiment is limited, mixed reality body doubling could be a promising form of exploration, especially with its high level of seamless transition between the physical and virtual worlds. 
These approaches could also reduce barriers for people in creating optimal environments with optimal conditions that would otherwise be difficult to achieve through live videos or through analog approaches such as a library \cite{Lee2021-rj}.

Furthermore, applying the Type of Task dimension highlights that there are opportunities to implement many different activities for the body double in mixed reality that one cannot can not as easily request on-demand in real-world setups \cite{Lee2021-rj}.  
Increasing the variety of tasks is especially important in making a body doubling setup customizable for an individual's needs in general, as well as in the exact moment. As discussed by Spiel et al. \cite{Spiel2022-ea}, adult life is less structured and therefore a body doubling system that adapts to changing conditions in everyday life is essential. 
We can envision endless possibilities in mixed reality, such as a setup with animal avatars with attentive awareness levels and medium levels of interaction in a café setting. Mixed reality environments could serve as a flexible and readily configurable testbed for customizing body doubling setups with various combinations of dimensions depicted in our model. 
Furthermore, its higher levels of immersion allow higher \textit{levels of awareness} and \textit{higher embodiment} of agents. A mixed reality setup has the potential of combining the realisim and immersion found in a real-life ``same time, same place'' with the customization typically found in 2D body double videos. A mixed reality version of LoFi Girl for example could incorporate the sounds and 3D visuals of each page turn, the calm presence of her cat, and a more immersive view from her window. 
\subsection{Extending Body Doubling}
Body doubling generally expects more parallel work and periods of time in which the individual engages in focused work. Deshmukh \cite{Deshmukh2025-pl} and Lalwani et al. \cite{Lalwani2025} explored possibilities of helpful interventions or interactive features that aim to support people with neurodiversity. In O'Connell et al.'s user study on a robot body double, they found that 73\% of their participants desired attention monitoring and distraction detection features \cite{OConnell2024}. These extensions would enable novel body doubling agents to reach higher levels of awareness and interaction that are at this time limited to human agents. A mixed reality or robot body double has the potential to detect moments of distraction or low motivation and could provide subtle cues or nudges or communicate directly. 


\subsection{Opportunities for Empirical Studies}
There is still a gap in terms of empirical studies on the mechanisms and efficacy of body doubling, in general or specifically for \ac{ADHD} \cite{Spiel2022-ea, Tan2025-lr}. As far as we are aware, existing research on body doubling and \ac{ADHD} is limited to only a small number of papers and master's theses \cite{Ara2025-bv, Lalwani2025, OConnell2024, born2024effects}. Further studies on different user groups with different attributes such as different forms of neurodiversity or different age groups would be beneficial in exploring the efficacy of different variables and dimensions as well as the overall construct of body doubling.  
\section{Conclusion}
Due to the limited amount of literature on body doubling in general, we developed a work-in-progress body doubling framework that can serve as a roadmap for further research in body doubling for adults with \ac{ADHD}. Based on existing literature, we defined dimensions that describe body doubling setups in terms of individual motivation, agent characteristics, interaction characteristics between agents, and contextual dimensions. The model may potentially help researchers brainstorm many possible mixed reality body doubling prototypes by varying the dimension values, envisioning more extended forms of body doubling, and using model's dimensions to construct empirical studies on body doubling. These various research directions may provide more concrete ideas on understanding and improving management systems for adults with \ac{ADHD}.

In general, the variety of different dimensions which we described in our body doubling framework show a massive diversity of existing and not-yet existing body doubling approaches, whose unique combination of characteristics could impact the efficacy of body doubling. Therefore, it is important that empirical research in body doubling takes into account of how different specific dimensions impact how effective body doubling is and ultimately whether it improves the quality of life of adults with \ac{ADHD}.






\section*{Declaration on Generative AI}
  The authors have not employed any Generative AI tools.

\bibliography{bibliography}

@ARTICLE{Sjowall2013-yo,
  title     = "Multiple deficits in {ADHD}: executive dysfunction, delay
               aversion, reaction time variability, and emotional deficits",
  author    = "Sj{\"o}wall, Douglas and Roth, Linda and Lindqvist, Sofia and
               Thorell, Lisa B",
  journal   = "J. Child Psychol. Psychiatry",
  publisher = "Wiley",
  volume    =  54,
  number    =  6,
  pages     = "619--627",
  month     =  jun,
  year      =  2013,
  copyright = "http://onlinelibrary.wiley.com/termsAndConditions\#vor",
  language  = "en"
}

@ARTICLE{Weibel2020-sa,
  title     = "Practical considerations for the evaluation and management of
               Attention Deficit Hyperactivity Disorder ({ADHD}) in adults",
  author    = "Weibel, S and Menard, O and Ionita, A and Boumendjel, M and
               Cabelguen, C and Kraemer, C and Micoulaud-Franchi, J-A and
               Bioulac, S and Perroud, N and Sauvaget, A and Carton, L and
               Gachet, M and Lopez, R",
  
  journal   = "Encephale",
  publisher = "Elsevier BV",
  volume    =  46,
  number    =  1,
  pages     = "30--40",
  month     =  feb,
  year      =  2020,
  language  = "en"
}

@misc{icd11_2025,
	title = {International {Classification} of {Diseases} 11th {Revision}. {The} global standard for diagnostic health information},
	url = {https://icd.who.int/en},
	urldate = {2026-02-12},
	author = {World Health Organization.},
	year = {2025},
}

@article{Holst2019,
  title = {Functional impairments among adults with {ADHD}: A comparison with adults with other psychiatric disorders and links to executive deficits},
  volume = {27},
  ISSN = {2327-9109},
  url = {http://dx.doi.org/10.1080/23279095.2018.1532429},
  DOI = {10.1080/23279095.2018.1532429},
  number = {3},
  journal = {Applied Neuropsychology: Adult},
  publisher = {Informa UK Limited},
  author = {Holst,  Ylva and Thorell,  Lisa B.},
  year = {2019},
  month = jan,
  pages = {243–255}
}

@ARTICLE{Tan2024-mr,
  title        = "Challenges in mixed reality in assisting adults with {ADHD}
                  symptoms",
  author       = "Tan, Valerie and Gerken, Jens",
  year         =  2024,
  primaryClass = "cs.HC",
  eprint       = "2508.07854"
}

@ARTICLE{Tan2025-lr,
  title         = "Preliminary results of a scoping review on assistive
                   technologies for adults with {ADHD}",
  author        = "Tan, Valerie and Jost, Luisa and Gerken, Jens and Pascher,
                   Max",
  month         =  oct,
  year          =  2025,
  copyright     = "http://creativecommons.org/licenses/by/4.0/",
  archivePrefix = "arXiv",
  primaryClass  = "cs.HC",
  eprint        = "2601.21791"
}

@inproceedings{OConnell2024,
author = {O'Connell, Amy and Banga, Ashveen and Ayissi, Jennifer and Yaminrafie, Nikki and Ko, Ellen and Le, Andrew and Cislowski, Bailey and Mataric, Maja},
title = {Design and Evaluation of a Socially Assistive Robot Schoolwork Companion for College Students with {ADHD}},
year = {2024},
isbn = {9798400703225},
publisher = {Association for Computing Machinery},
address = {New York, NY, USA},
url = {https://doi.org/10.1145/3610977.3634929},
doi = {10.1145/3610977.3634929},
booktitle = {Proceedings of the 2024 ACM/IEEE International Conference on Human-Robot Interaction},
pages = {533–541},
numpages = {9},
location = {Boulder, CO, USA},
series = {HRI '24}
}

@ARTICLE{Ginsberg2014-pn,
  title    = "Underdiagnosis of attention-deficit/hyperactivity disorder in
              adult patients: a review of the literature",
  author   = "Ginsberg, Ylva and Quintero, Javier and Anand, Ernie and
              Casillas, Marta and Upadhyaya, Himanshu P",
  journal  = "Prim. Care Companion CNS Disord.",
  volume   =  16,
  number   =  3,
  month    =  jun,
  year     =  2014,
  language = "en"
}

@INPROCEEDINGS{Spiel2022-ea,
  title      = "{ADHD} and technology research -- investigated by
                neurodivergent readers",
  booktitle  = "{CHI} Conference on Human Factors in Computing Systems",
  author     = "Spiel, Katta and Hornecker, Eva and Williams, Rua Mae and Good,
                Judith",
  publisher  = "ACM",
  month      =  apr,
  year       =  2022,
  address    = "New York, NY, USA",
  conference = "CHI '22: CHI Conference on Human Factors in Computing Systems",
  location   = "New Orleans LA USA"
}

@ARTICLE{Eagle2024-cb,
  title     = "``{I}t was something {I} naturally found worked and heard about
               later'': An Investigation of Body Doubling with Neurodivergent
               Participants",
  author    = "Eagle, Tessa and Baltaxe-Admony, Leya Breanna and Ringland,
               Kathryn E",
  journal   = "ACM Trans. Access. Comput.",
  publisher = "Association for Computing Machinery (ACM)",
  month     =  aug,
  year      =  2024,
  language  = "en"
}

@INPROCEEDINGS{Eagle2023-tm,
  title      = "Proposing body doubling as a continuum of space/time and
                mutuality: An investigation with neurodivergent participants",
  booktitle  = "The 25th International {ACM} {SIGACCESS} Conference on
                Computers and Accessibility",
  author     = "Eagle, Tessa and Baltaxe-Admony, Leya Breanna and Ringland,
                Kathryn E",
  publisher  = "ACM",
  pages      = "1--4",
  month      =  oct,
  year       =  2023,
  address    = "New York, NY, USA",
  copyright  = "https://www.acm.org/publications/policies/copyright\_policy\#Background",
  conference = "ASSETS '23: The 25th International ACM SIGACCESS Conference on
                Computers and Accessibility",
  location   = "New York NY USA"
}

@ARTICLE{Taber2019-ad,
  title     = "What makes a live stream companion?",
  author    = "Taber, Lee and Baltaxe-Admony, Leya Breanna and Weatherwax,
               Kevin",
  journal   = "Interactions",
  publisher = "Association for Computing Machinery (ACM)",
  volume    =  27,
  number    =  1,
  pages     = "52--57",
  month     =  dec,
  year      =  2019,
  copyright = "http://www.acm.org/publications/policies/copyright\_policy\#Background",
  language  = "en"
}

@ARTICLE{Ara2025-bv,
  title        = "You are not alone: Designing body doubling for {ADHD} in
                  virtual reality",
  author       = "Ara, Zinat and Rahim, Imtiaz Bin and Zhou, Puqi and Yu,
                  Liuchuan and Esmaeili, Behzad and Yu, Lap-Fai and Hong,
                  Sungsoo Ray",
  year         =  2025,
  primaryClass = "cs.HC",
  eprint       = "2509.12153"
}

@INPROCEEDINGS{Lee2021-rj,
  title      = "Personalizing ambience and illusionary presence: How people use
                ``study with me'' videos to create effective studying
                environments",
  booktitle  = "Proceedings of the 2021 {CHI} Conference on Human Factors in
                Computing Systems",
  author     = "Lee, Yoonjoo and Chung, John Joon Young and Song, Jean Y and
                Chang, Minsuk and Kim, Juho",
  publisher  = "ACM",
  month      =  may,
  year       =  2021,
  address    = "New York, NY, USA",
  conference = "CHI '21: CHI Conference on Human Factors in Computing Systems",
  location   = "Yokohama Japan"
}

@INPROCEEDINGS{Deshmukh2025-pl,
  title      = "Toward neurodivergent-aware productivity: A systems
                {and AI-based} human-in-the-loop framework for 
                {ADHD-affected professionals}",
  booktitle  = "Proceedings of the 16th Biannual Conference of the Italian
                {SIGCHI} Chapter",
  author     = "Deshmukh, Raghavendra",
  publisher  = "ACM",
  pages      = "1--6",
  month      =  oct,
  year       =  2025,
  address    = "New York, NY, USA",
  conference = "CHItaly 2025: CHItaly 2025: 16th Biannual Conference of the
                Italian SIGCHI Chapter",
  location   = "Salerno Italy"
}

@inproceedings{Lalwani2025,
author = {Lalwani, Himanshi and Saleh, Mira and Salam, Hanan},
title = {A Study Companion for Productivity: Exploring the Role of a Social Robot for College Students with {ADHD}},
year = {2025},
publisher = {IEEE Press},
booktitle = {Proceedings of the 2025 ACM/IEEE International Conference on Human-Robot Interaction},
pages = {1438–1442},
numpages = {5},
location = {Melbourne, Australia},
series = {HRI '25}
}

@masterthesis{annavarapu2024comparative,
  title={Comparative Study of Body Doubling in Extended Reality},
  author={Annavarapu, Swetha},
  year={2024},
  school={Virginia Tech}
}

@mastersthesis{born2024effects,
  title={The Effects of Accountability and Body Doubling on Productivity for People With Attention Deficits},
  author={Born, Sydnie},
  year={2024},
  school={Lamar University-Beaumont}
}

@misc{weisstein_completedigraph,
  author       = {Weisstein, Eric W.},
  title        = {Complete {D}igraph},
  howpublished = {From {MathWorld}---A {W}olfram {W}eb {R}esource},
  url          = {https://mathworld.wolfram.com/CompleteDigraph.html},
  year = {2026},
  note         = {Accessed: 2026-02-10}
}

@InProceedings{Jhuang_2022,
author="Jhuang, Yi-Ci
and Chiu, Yu Hsien
and Lee, Hsuan-Jen
and Lee, Yen Ting
and Lin, Guan-You
and Wu, Nien-Hsin
and Kuo, Pei-Yi Patricia",
editor="Stephanidis, Constantine
and Antona, Margherita
and Ntoa, Stavroula",
title="Exploring the Effect of Study with Me on Parasocial Interaction and Learning Productivity: Lessons Learned in a Field Study",
booktitle="HCI International 2022 Posters",
year="2022",
publisher="Springer International Publishing",
address="Cham",
pages="43--49",
isbn="978-3-031-06391-6"
}

@misc{bodydoublingDiscordBody,
	author = {},
	title = {{D}iscord {F}{A}{Q} - {B}ody {D}oubling --- bodydoubling.com},
	howpublished = {\url{https://bodydoubling.com/discord-faq/}},
	year = {2025},
	note = {[Accessed 27-05-2025]},
}

@ARTICLE{Gutwin2002-by,
  title     = "A descriptive framework of workspace awareness for real-time
               groupware",
  author    = "Gutwin, Carl and Greenberg, Saul",
  journal   = "Comput. Support. Coop. Work",
  publisher = "Springer Science and Business Media LLC",
  volume    =  11,
  number    = "3-4",
  pages     = "411--446",
  month     =  sep,
  year      =  2002,
  copyright = "https://www.springernature.com/gp/researchers/text-and-data-mining",
  language  = "en"
}

@article{PENICHET2007237,
title = {A Classification Method for {CSCW} Systems},
journal = {Electronic Notes in Theoretical Computer Science},
volume = {168},
pages = {237-247},
year = {2007},
note = {Proceedings of the Second International Workshop on Views on Designing Complex Architectures (VODCA 2006)},
issn = {1571-0661},
doi = {https://doi.org/10.1016/j.entcs.2006.12.007},
url = {https://www.sciencedirect.com/science/article/pii/S1571066107000394},
author = {V.M.R. Penichet and I. Marin and J.A. Gallud and M.D. Lozano and R. Tesoriero}
}

@book{johansen1988groupware,
  title={Groupware: Computer Support for Business Teams},
  author={Johansen, Robert and Charles, Jeff and Mittman, Robert and Saffo, Paul},
  year={1988},
  publisher={Free Press},
  address={New York},
  isbn={0-02-916491-5}
}

@misc{addADHDBody,
	author = {ADDA Editorial Team},
	title = {{T}he {A}{D}{H}{D} {B}ody {D}ouble: {A} {U}nique {T}ool for {G}etting {T}hings {D}one - {A}{D}{D}{A} - {A}ttention {D}eficit {D}isorder {A}ssociation --- add.org},
	howpublished = {\url{https://add.org/the-body-double/}},
	year = {2024},
	note = {[Accessed 27-05-2025]},
}

@online{lofigirl_channel,
  author = {{Lofi Girl}},
  title = {Lofi Girl - {YouTube} Channel},
  year = {2026},
  url = {https://www.youtube.com/@LofiGirl},
  note = {Accessed: 2026-02-20}
}

@ARTICLE{Gjervan2012-xe,
  title     = "Functional impairment and occupational outcome in adults with
               {ADHD}",
  author    = "Gjervan, Bj{\o}rn and Torgersen, Terje and Nordahl, Hans M and
               Rasmussen, Kirsten",
  journal   = "J. Atten. Disord.",
  publisher = "SAGE Publications",
  volume    =  16,
  number    =  7,
  pages     = "544--552",
  month     =  oct,
  year      =  2012,
  language  = "en"
}

@ARTICLE{Moell2015-az,
  title     = "Living {SMART} --- A randomized controlled trial of a guided
               online course teaching adults with {ADHD} or sub-clinical {ADHD}
               to use smartphones to structure their everyday life",
  author    = "Mo{\"e}ll, Birger and Kollberg, Linn{\'e}a and Nasri, Berkeh and
               Lindefors, Nils and Kaldo, Viktor",
  journal   = "Internet Interv.",
  publisher = "Elsevier BV",
  volume    =  2,
  number    =  1,
  pages     = "24--31",
  month     =  mar,
  year      =  2015,
  copyright = "http://creativecommons.org/licenses/by-nc-nd/3.0/",
  survey = "•"
}

\end{document}